\documentclass[12pt,preprint]{aastex}

\shorttitle{Electric dipole emission by fullerenes and buckyonions} 
\shortauthors{Susana Iglesias-Groth}  

\begin{document}

\title{Electric dipole emission by fulleranes and Galactic anomalous microwave emission} 
\author{Susana Iglesias-Groth}  
\affil{Instituto de Astrof\'\i sica de Canarias, C/ Via L\'actea s/n, 38200 La
Laguna, Spain}

\begin{abstract}

We study the rotation rates and  electric dipole emission of hydrogenated 
icosahedral fullerenes (single and multishell)  in various phases of the interstellar medium. 
Using the formalism of Draine and Lazarian for the rotational dynamics of these
 molecules in various astrophysical environments, we find  effective rotation rates 
in the range 1--65 GHz with a trend toward lower rotational frequency as the radius of 
the molecule increases.   Owing to the moderately polar nature of the C--H bond, 
hydrogenated fullerenes (fulleranes)  are expected to have a net dipole moment
 and produce  electric dipole radiation. Adopting   the same
 size distribution proposed for fullerenes  in the study of the UV extinction bump (2175 \AA)
  we predict the dipole electric emission of mixtures of  fulleranes for
 various levels of hydrogenation. We find that these molecules  could be the carriers
 of  the anomalous microwave emission recently detected by Watson et al.\ in the Perseus
 molecular complex.

\end{abstract} 
 
\section{Introduction}

 Recent experiments dedicated to the study of the anisotropy of the cosmic 
microwave background have found evidence for galactic  microwave emission in the range 10--90 GHz 
correlated at high galactic latitudes with thermal emission (DIRBE 100 $\mu$m  
map) from interstellar dust (Kogut et al.\ 1996; Leitch et al.\ 1997; 
de Oliveira-Costa et al.\ 1999, 2002, 2004). An explanation for this anomalous dust-correlated 
microwave emission based on electric dipole emission from fast rotating carbon-based  
molecules has been proposed by Draine \& Lazarian 1998a,b). These  models 
appear to reproduce the major features of the so-called anomalous microwave emission
(Finkbeiner et al.\ 2004) but do not identify the actual carrier of the emission.

 Iglesias-Groth (2004) has recently shown  that photoabsorption by fullerenes and buckyonions
 (multishell fullerenes) in the interstellar medium (ISM) can account for the UV bump (217.5 nm) of 
the interstellar extinction curve and  possibly for some of the diffuse interstellar bands.
It is suggested that  fullerenes and buckyonions are broadly distributed in the Galaxy and
 contain a  large  fraction of interstellar carbon.  
The intrinsic dipole moment of these  highly symmetric  molecules is rather small, thus  it is 
unlikely that they  contribute significantly to the anomalous microwave emission. However,
it is expected that a fraction of fullerenes in the interstellar medium will be hydrogenated;
the so-called fulleranes would have a much higher dipole moment because of the polar nature
of the C-H bond. Experimental results on the electric dipole moment of isolated  alkali-C$_{60}$ molecules show an 
increase from 12.4 D (Debye) for C$_{60}$Li to 
21.5 D for C$_{60}$Cs and indicate a strong electron transfer from the alkali atom to the C$_{60}$
cage that is almost complete for the  largest alkalis (Antoine et al.\ 2000). Given this
behavior it is reasonable to expect that the dipole moment of  fulleranes is higher than
tha of the  C-H bond (0.3 D) and therefore, these molecules are potential carriers 
of electric dipole emission.

Fulleranes have deserved  attention as
 potential carriers of diffuse intestellar bands and other interstellar and circumstellar
 features (Webster 1991, 1992,1993a,b). Both, fullerenes and fulleranes  have also  been
 detected in  samples of the  Allende meteorite (Becker \& Bunch 1997) which suggests
 their existence in the interstellar medium. 
  In this paper, we study the electric dipole emission of fulleranes with various level of hydrogenation 
  in representative physical conditions in the  interstellar medium and show their potential as carriers of the anomalous 
  microwave emission in the Galaxy. The predicted
emissivities are compared with very recent measurements of anomalous microwave emission in the Perseus
molecular complex (Watson et al.\ 2005).

\section{BASIC MOLECULAR PROPERTIES: ELECTRIC DIPOLE  AND INERTIA MOMENTS.}

%\subsection{Radii, inertia moment and dipole moments for  fullerenes and  buckyonions}

We will consider here the various families of fulleranes C$_N$ H$_p$  that can be formed from icosahedral
 fullerenes with $N$ = 60, 180, 240, 540, 960, 1500, and 2160 carbon atoms.
Fullerenes can also adopt multilayered configurations in which one is encapsulated inside another 
with separation of $\sim$ 3.4--3.5 \AA.
 These multishell fullerenes, also called buckyonions, can also 
 be hydrogenated. We will consider here only those buckyonions with a complete number of shells
 and with  hydrogen atoms bonded only in the outermost  shell. 
 At most, one hydrogen atom may be bound to each of the carbon atoms of a fullerene;
thus, the number, $P$, of  hydrogen atoms  of any fullerane or buckyonion  will lie in the range  $P=1$ to $N$. 
The ratio between the number of hydrogen and carbon atoms will be, $s=P/N$.
We will deal with the following types of  chemical species: C$_{N}$H$_{P}$, C$_{N}^{-}$H$_{P}$, 
C$_{N}^{+}$H$_{P}$, C$_{N}$H$^{+}$, and the corresponding hydrogenated buckyonion analogs  
(C$_{60}@$C$_{240}@...$C$_N $H$_P$, etc.).  Fullerene radii, $R_C$, were adopted as in Iglesias-Groth et al.\  (2002). They 
range between 3.55 \AA~ and 21.5 \AA~ for C$_{60}$ and   C$_{2160}$, respectively. The hydrogen atoms
of the fulleranes are expected to be located radially outward from the carbon atoms.
The length of the C--H bond will be taken as in benzene 1.07 \AA~ (see, for example, Braga et al.\ 1991).
We assume  the  moment of inertia, $I_C$,  of a fullerene  as  that of a spherical cage with radius
$ $R    :  $I_C= 2/3 N m_c R^2$ , where  $m_c$ is the mass of the carbon atom. For the ensemble
of fulleranes with $P$ hydrogen atoms we adopt as moment of inertia the sum of the moment of inertia of the 
 relevant fullerenes, I$_C$,  with the moment of inertia of an hypothetical  spherical cage of mass 
$P$ times the mass of the hydrogen atom, and radius  $R+1.07$ \AA. 
The moment of inertia of the hydrogenated buckyonion will be taken as  the sum of the 
moments of inertia of all the  individual fullerenes conforming the molecule plus 
that of the layer of hydrogen atoms.

We will assume in what follows that  the dipole moment of a fullerane (or hydrogenated
buckyonion) is proportional to 
the dipole moment of the C--H bond ($\kappa$0.3 D) and that the hydrogen atoms are randomly
 distributed on the surface of the cage in such a way that  the intrinsic dipole moment of the  
fullerane with $P$ hydrogen atoms can be approximated by $\mu\sim \kappa~ 0.3 P^{1/2}$. We will adopt 
 a value $\kappa=1$ in the calculations.
  Owing to collisions with ions and electrons and also because of photoelectric emission,  
fullerenes in interstellar space may be charged.
An additional electric dipole component could be expected as a
 consequence of any  displacement between the centroid of the molecule charge 
 and the center of mass (Purcell 1979).   Given the high
 symmetry of our molecules we will assume these displacements produce negligible additional
dipole moments and  will thus adopt the same value for the  dipole moments of 
  C$_N^{+}$H$_P$, C$_N^{-}$H$_P$ and  C$_N$H$_P$. Fulleranes  of the form C$_N$ H$_P^+$, where a proton has been adsorbed instead of a 
neutral hydrogen atom, will significantly increase  their intrinsic dipole moment. Given
the absence of experimental information, we have estimated what would be the permanent
dipole moment for this type of molecule and also for protonated  buckyonions,
 using the formalism in Rittner (1951). This takes into account the relatively
 high  polarization contribution of the fullerene cage. 
Values for the  polarizabilities of the fullerenes and buckyonions have been adopted from 
the semiempirical model of  Iglesias-Groth et al.\ (2002, 2003). 
In particular, this formalism gives 5.9 D for the dipole moment of  C$_{60}$H$^{+}$.
 Since this formalism may overestimate the dipole moment of these molecules we will
conservatively assume  in these cases that the  value of the dipole moment is  that of the 
 C--H$^+$ bond (1.68 D, Follmeg et al.\ 1987).

\section{ROTATION AND  ELECTRIC DIPOLE EMISSIVITY FROM FULLERANES.}

To study the rotational damping
and excitation mechanisms affecting fulleranes and hydrogenated buckyonions in the ISM we will use
 the formalism
developed by Draine \& Lazarian (1998a,b). We will consider two representative environments 
in the interstellar medium: a)  The so-called cold neutral medium (CNM) characterized by  
low hydrogen density
of $n_{\rm H}$= 30  cm$^{-3}$ , gas temperature of $T=100$ K, and low ionization fraction $n$(H$^+$)$/$n$_{\rm H}=0.0012$;  and 
b) the warm ionized medium (WIM) with very low density  $n_{\rm H}$ = 0.1  cm$^{-3}$ , 
gas temperature of $T=8000$ K, and  high ionization fraction $n$(H$^+$)/$n_{\rm H}$ = 0.99. In thermal equilibrium, fulleranes  share the same temperature as the gas 
and estimation of the  rms rotation frequency is straightforward.
When the temperature of  fulleranes and  gas are different,  some processes will produce
rotational damping  and others will lead to rotational excitation. Draine and Lazarian discuss 
 both types of processes comprehensively. Following their treatment, we compute the rotational damping produced by 
{\itshape collisional drag}, associated  with H, H$_{2}$, and He species (neutral and ions) temporarily stuck 
and subsequently desorbed from  the surface of the fullerenes;  {\itshape plasma drag} caused by the 
 the interaction  of the electric dipole moment, $\mu$, of the fullerene and the 
electric dipole of the passing ions;    and the {\itshape infrared emission},  due to the thermal emission of 
photons that previously heated the molecule.
We also compute rotation excitation rates associated with the impact of neutral
particles and  ions,  plasma drag, and infrared emission.

 We have assumed  the temperature of the fulleranes  is 20 K in the CNM and WIM
phases of the ISM. In Figure 1 we show  results for the  effective rotation rates,
\begin{equation}
\omega_{\rm rad}=(\frac{5}{3})^{0.25}<\omega^2>^{0.5},
\end{equation}
of fulleranes with different degree of hydrogenation. We illustrate the case for the CNM phase of the ISM. The rotation rate 
decreases as the radius of the molecule increases. Buckyonions show slower rotation rates due to the higher moment of inertia. 
Increasing the level of hydrogenation for molecules of a given radius (increasing the dipole moment) leads to lower rotation rates. 
We display results for three levels of hydrogenation $s=1/N$, 1/10, and 1/3,  where $N$ is the number of carbon atoms in the fullerane or 
in the most external shell of the buckyonion. Assuming $\kappa=1$, in the case of C$_{60}$ these hydrogenation ratios correspond to 
dipole moments of 0.3, 0.74, and 1.34 D, respectively; and 0.3, 4.4, and 8.1 D for C$_{2160}$, the largest fullerane in this study. 
Hydrogenation ratios as high as $s=1/3$ are not rare among other carbon based molecules, for example PAHs, and  as discussed by 
Webster (1992) may be rather common in fulleranes. For comparison, we also display the rotation rates of singly protonated 
fullerenes in WIM conditions. We find lower rotation rates than in the CNM case.

Emissivity curves for individual fulleranes and buckyonions of various sizes are plotted in Figure 2a for CNM conditions. For 
these computations, we  assume that the densities of C$_{60}$H$_p$ and C$_{180}$H$_p$ are 0.2 $\times$ 10$^{-6}$ $n_{\rm H}$,
 as obtained for 
similar size fullerenes  in  Iglesias-Groth (2004). For the larger fulleranes  abundances are scaled down  according to the 
size distribution law $n(R)\varpropto R^{-m}$, with $m$ =3.5 $\pm$ 1.0 proposed in the previous paper. We are assuming $\kappa=1$ 
and illustrate results  for the case $s=1/3$.  We find that for a given level  of hydrogenation the major emitters are the 
smallest fulleranes, and that the frequency of peak emission decreases as the radius increases.  We also plot the total 
emissivity curve  for  a mixture of monoshell fulleranes following  the proposed size distribution law. The 
computations have been  extended  up to fulleranes with N=2160, but  clearly the largest fulleranes have negligible 
contribution to the total emissivity curve. Remarkably, this curve shows significant emission (above 10\% of the peak)  
in the range 10 to 65 GHz with  a peak near 25 GHz.  For comparison, we also plot emissivity curves for  a mixture of 
hydrogenated (same $s$ value) icosahedral   buckyonions with a complete number of shells, starting with   C$_{60}@$C$_{240}$H$_p$. 
We assume the same abundance for this molecule as for the single C$_{240}$H$_p$. The emissivity of these  buckyonions is similar 
to  that of the indilvidual fullerane of same radius. We also plot the emissivity curve that would correspond to a mixture of
 hydrogenated buckyonions following the previous size distribution. In this case, most of the emission lies in the frequency 
 range 10--30 GHz with a peak at $\sim$18 GHz.  For the same level of hydrogenation, the  buckyonion mixture has a much lower 
 emissivity than the mixture of  single-shell fulleranes.

In panel b of Figure 2 we show how the predictions are sensitive to the level of hydrogenation and to the physical conditions of
the medium. We present total emissivity curves for a mixture of fulleranes and hydrogenated buckyonions (following the size distribution 
of index, $m=-3.5$) with three different levels of  
hydrogenation ($s=1/N$, 1/10, and 1/3) for both CNM  and WIM conditions. Increasing the level of hydrogenation (i.e., 
increasing  the average dipole moment of the molecules) leads to a narrower frequency range for the emission. The abundances of the  smaller fullerenes (C$_{60}$ and C$_{180}$) have 
been assumed the same $0.2 \times 10^{-6}$ $n_{\rm H}$.  WIM emissivity curves are found narrower and shifted to lower frequencies. 
 Under WIM conditions it would be
plausible to expect some degree of protonization of the fulleranes. As an indication of how significant this could be regarding 
electric dipole emission, we show
in the figure the emissivity curve obtained for  the case $s=1/N$, but considering  a proton instead of a hydrogen atom. 
In this case, the much larger dipole moment of the molecules  makes the peak emission to shift down almost  a
 factor 2.
In summary, under CNM conditions we expect the hydrogenated fullerene/buckyonion mixture produces electric dipole emission with a 
peak in the range 20--50 GHz which shifts to lower frequencies as increases the hydrogenation level. In WIM conditions, this emission 
tends to be
narrower and shifted to lower frequency.

 Very recently, using data from the COSMOSOMAS experiment (11--17 GHz) in combination with  WMAP data, Watson et al. (2005) show  that on scales of a few degrees,  the   anomalous microwave 
emission detected towards the Perseus molecular complex dominates any other known galactic  emission process (synchotron, free--free, or dust emission) in the frequency range 10--50 GHz. The anomalous emission peaks at 22 GHz (with an integrated
flux of $\sim$40 Jy) very close to the peak 
emission that we have found for a mixture of fullerenes under CNM conditions. Watson et al.\ show how a mixture
of CNM and WIM of Draine and Lazarian models of electric dipole emission by spinning dust particles
may explain the results. The nature of the particles in the Draine and Lazarian 
models are very general, so it is interesting to  study  whether the anomalous microwave emission
 can be explained by fulleranes with a minimum
number of hypotheses. 

In Figure 3 we plot the empirical results obtained by Watson et al.\ (2005) and emissivity curves for mixtures of fulleranes and 
buckyonions computed as described above.
 The measurements are corrected following the prescriptions in Watson et al.\ for the
contribution of vibrational dust ($T_{\rm dust}=19$ K and emissivity index 1.55) and for the contribution of 
free-free in the Perseus region.  We  consider all the  measurements available in the frequency range where anomalous
emission appears to make a major contribution, i.e.\ from 10 to 90 GHz and overplot emissivity curves computed for
several minimal cases either using CNM conditions or a combination of CNM and WIM. We plot a few examples that 
show how the predicted emissivities are indeed very close to those observed.   Fitting  the
observations require some fine tuning between the number of molecules under CNM and WIM conditions and 
the level of hydrogenation. For simplicity, 
 we have assumed that in WIM conditions the fullerenes are singly protonated. First,
it is  apparent that a significant level
of hydrogenation is required if fulleranes are to explain the observed emission. Increasing the level of hydrogenation leads to a higher
 dipole moment and shifts the bulk of emission to lower frequencies.  However, 
the tail emission at high frequencies ($\gtrsim$ 40 GHz) also requires a contribution of fulleranes with a small dipole moment. 
Molecules under WIM conditions alone would not be able to explain this tail.
The dashed-dotted curve in Figure 3 corresponds to the emissivity of a mixture of  fulleranes and buckyonions in CNM,  where any possible 
degree of hydrogenation  has equal  weight. The curve has been scaled to 
 match approximately the measurement at 22 GHz, but the shape is inconsistent with the observations. The dotted line present the results 
 for  a mixture 
 assuming a range of  hydrogenation  peaking at  $s=1/10$, such that 40\% of the molecules of a given radius have $s=1/10$, 30\%  have a 
 factor 2 higher dipole moment and the remaining 30\% a factor 2 lower. The emissivity curves obtained 
for other  hydrogenation ranges do not give significantly  better
fits.  The experimental data are best reproduced (solid line)
 for a combination (with equal weight)
 of the previous curve and  the emissivity curve of singly protonated fullerenes in WIM conditions, again scaled  
to match the measurement at 22 GHz. This factor is close to 3 and results from a possible underestimation of
the dipole moments and/or number density of fulleranes in the region.  The dashed line
illustrates the same combination between CNM and WIM but modifying the  hydrogenation range of the molecules in CNM, which are now
 assumed to be uniformly distributed among  $s=1/10$, 1/20, and 1/30. 

  In summary,  it is possible to understand  the Perseus microwave anomalous emission
in terms of electric dipole emission of fulleranes if they follow a  size distribution similar
to that proposed in the study of the UV extinction bump. The dominant microwave emission would be  associated
to the smaller fullerenes. Lazarian and 
Draine (2000) have suggested that fast rotating ultrasmall ($<$ 10 \AA) grains may be aligned with the interstellar magnetic field via
 paramagnetic resonance relaxation and produce polarized electic dipole emission.  Recent reports 
that carbon based material like graphite (Esquinazzi et al. 2002) polymerized fullerenes 
(Makarova et al. 2001) and C$_{60}$H$_{24}$ hydrofullerites (Antonov et al. 2002)   exhibit a weak ferromagnetic moment 
 suggests that the magnetic properties  of  fulleranes could also lead to polarized rotational electric dipole emission.
  Precise measurements  of the intensity and polarization of the anomalous microwave emission and its relation with UV extinction in
 the Perseus molecular complex and other similar regions will be essential to establish the role of these molecules.

\section{CONCLUSIONS}
 We have computed effective rotation rates of hydrogenated fullerenes  in the ISM and find values in the  1--65 GHz range with a  clear trend toward lower rotation as the radius of 
the molecule increases. Increasing the level of hydrogenation shifts the peak of electric dipole emission to lower frequencies. We predict emissivity curves for fullerane mixtures assuming the abundance and  size
distribution $N(R) \propto R^{-3.5}$ proposed for fullerenes in the ISM. We find that
the dominant dipole electric emission would be due to fulleranes
with  radii smaller than $\sim$ 10--15 \AA~  and  dipole moments of order $\sim$ 1 D.
Changes in the degree of hydrogenation moves  the peak of the emissivity curves in the range 10--60 GHz. 
 The resulting curves for mixtures of fulleranes in CNM and WIM conditions reproduce the recent observations of anomalous 
 microwave emission in the Perseus molecular complex. The bulk of this emission can be explained  in terms of  electric dipole 
 radiation  of  small icosahedral fullerenes (C$_{60}$ to C$_{240}$) with a modest hydrogenation level. 
  Laboratory  measurements of the dipole moment of fulleranes  are crucial to the further study of these molecules as potential 
    carriers of the dust-correlated microwave emission detected by cosmic microwave background experiments.

\acknowledgments
I thank Rafael Rebolo and Bob Watson for valuable comments and suggestions on this work. I also thank the COSMOSOMAS team for early 
access to their results.

\begin{figure}
\includegraphics[angle=0,scale=.80]{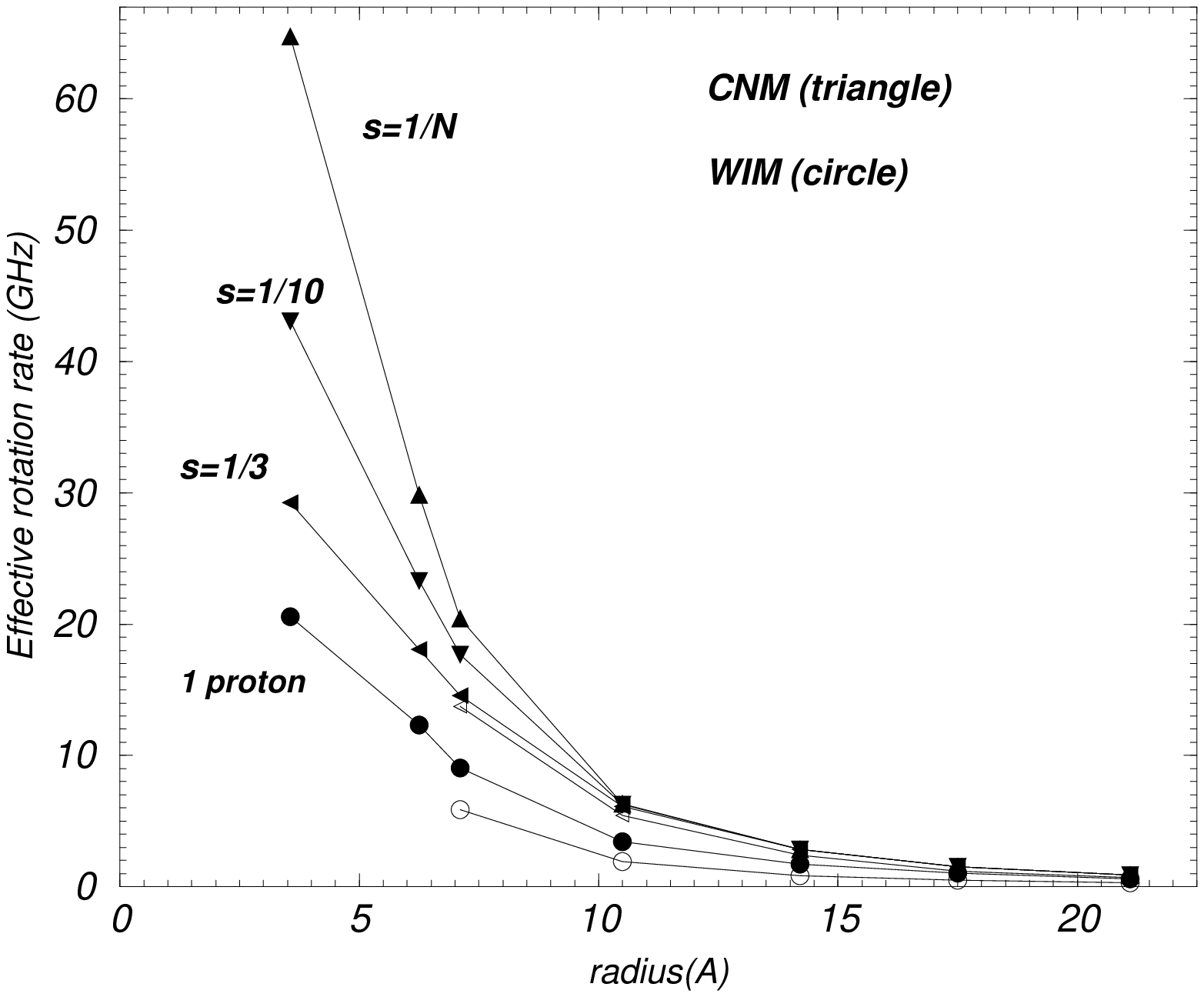}
\caption[]{
Effective rotation rates,~$\omega_{\rm rad}/2\pi$, as a function of radius for neutral fulleranes with various levels of 
hydrogenation (filled triangles), hydrogenated buckyonions (open triangles) in CNM conditions. The case of a fullerenes 
and buckyonions with one proton is shown as filled circles or open circles respectively.}
\end{figure}

\begin{figure}
\includegraphics[angle=0,scale=.80]{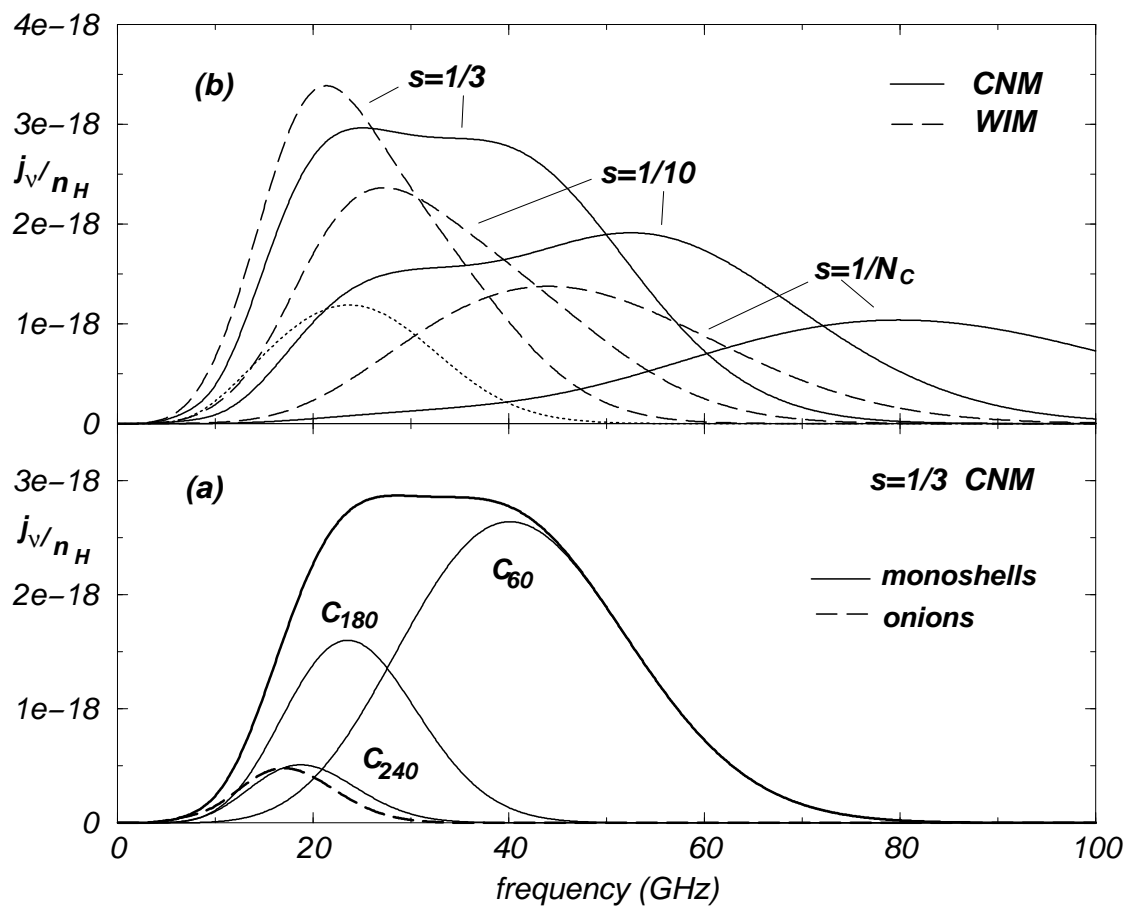}

\caption[]{a) Electric dipole emissivity per H (Jy cm$^2$ sr$^{-1}$ H$^{-1}$) of fulleranes (solid line)and hydrogenated buckyonions 
(dashed line) with hydrogenation ratio $s=1/3$ in CNM conditions. Results are plotted for C$_{60}$, C$_{180}$, and C$_{240}$,  and 
for a mixture of fulleranes and buckyonions. b) Electric dipole emissivity curves for a mixture of fulleranes and hydrogenated 
buckyonions with three different levels of hydrogenation for both CNM (solid line)  and WIM (dashed line) conditions. The dotted 
line is the emissivity due to a mixture of singly protonated fullerenes and buckyonions for WIM.}
\end{figure} 

\begin{figure}
\includegraphics[angle=0,scale=.80]{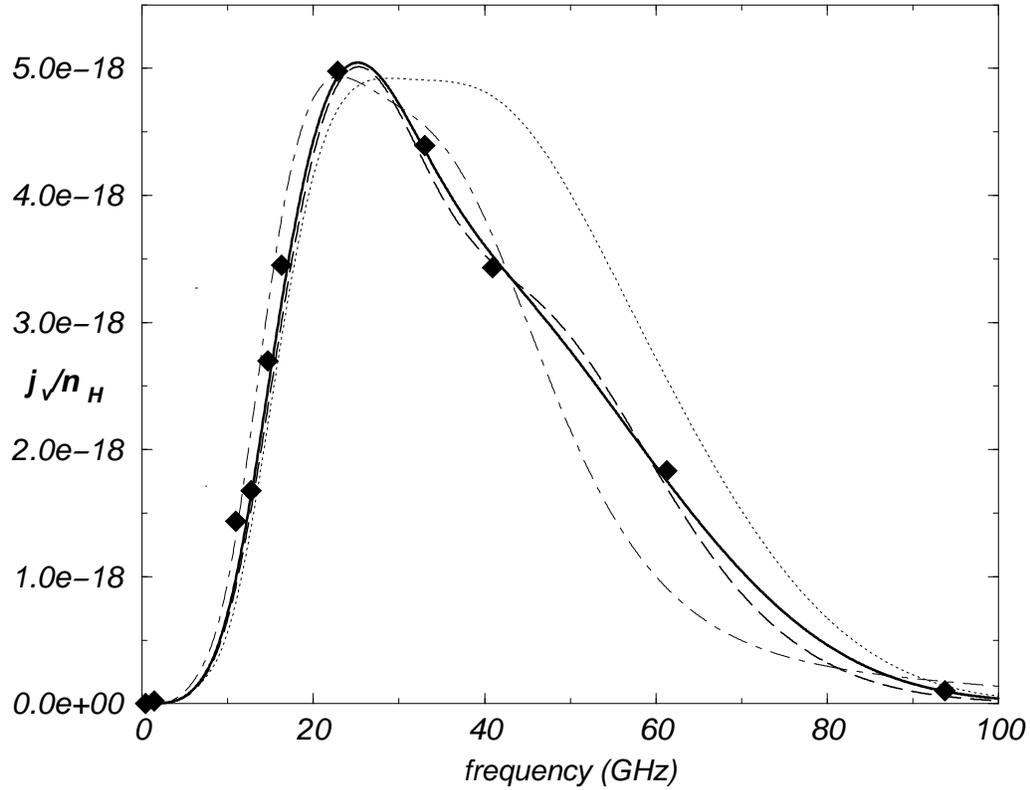}

\caption[]{Observations of Perseus anomalous microwave emission by Watson et al.\ 2005 (filled diamonds) and predicted 
rotational emissivity per H (Jy cm$^2$ sr$^{-1}$ H$^{-1}$) of a mixture of fulleranes and hydrogenated buckyonions in CNM conditions 
(dotted and dotted-dashed line) and a combination of  CNM and WIM conditions (dashed line and solid line). See text for details.}

\end{figure} 

\end{document}